\documentclass[prd,aps,twocolumn]{revtex4}
\begin{document}
\title{A simple mechanical analog of the field theory of tachyon matter}
\author{Sayan Kar \footnote{Electronic address: {\em sayan@cts.iitkgp.ernet.in}}
${}^{}$
}
\address{Department of Physics and Centre for
Theoretical Studies \\Indian Institute of Technology, Kharagpur 721 302, India}
\begin{abstract}
In this brief note we show that the zero dimensional version of the field 
theory of tachyon matter, proposed by Sen, 
provides an action integral formulation for the 
motion of a particle in the presence of Newtonian gravity 
and nonlinear damping (quadratic in velocity).   
\end{abstract}
\maketitle


\maketitle
\vspace{.2in}

Recently, there has been a lot of interest around a proposal
by Sen {\cite{sen1}}, of a new field theory  
--- the field theory of tachyon matter. The derivation of the
action is based on involved arguments in string and
string field theory {\cite{sen1,sen2}}. This action has also been
proposed earlier, independently, by several authors in
{\cite{others}}, {\cite{panda}}, though, it is only recently 
that it has been analysed in greater detail{\cite{sen1,sen2}}.
Apart from its importance in the context of string theory,
it is useful to
investigate its consequences by considering it 
as a field theory in its own right. 
It goes without saying, that, at first sight, one
is somewhat surprised by the expression for the action, given as :

\begin{equation}
S = -\int d^{p+1}x V(T)\sqrt{1+\eta^{ij}\partial_i T\partial_j T}
\end{equation}

\noindent where $\eta_{00} = -1 $ and $\eta_{\alpha\beta} = \delta_{\alpha\beta}$
with $\alpha$, $\beta = 1,2,...p$,
$T(x)$ is the scalar tachyon field and  V(T) is the tachyon potential,
which, from string field theory arguments is found to be given by : 

\begin{equation}
V(T) \sim e^{-\alpha T/2} 
\end{equation}

\noindent where $ \alpha=1$ (bosonic, $p=25$) and ,$\alpha =  \sqrt{2}$
(superstring, $p=9$).

The tachyon potential appears, in the action, as a multiplicative factor to 
a term with a square root. This
is what makes this theory unfamiliar in comparison to usual field theories
known to us.
This field theory has been analysed extensively in recent papers by
Sen {\cite{sen2}}. There has been a flurry of recent work on the cosmological
relevance/implications of tachyonic matter {\cite{gibbons}} motivated by the 
fact that
the effective energy--momentum tensor is equivalent to that of
noninteracting, nonrotating dust {\cite{sen1}.

At a pedagogical level, it is certainly true, to some extent atleast,
that lower dimensional analogs
of a field theory helps our understanding in some way, though
working with such correspondences may sometimes be entirely useless
( a prime example is lower dimensional gravity which doesn't seem
to have much connection with actual 4D gravity despite the volumes that
have been written on it). 
It also provides an useful method of
introducing the theory to undergraduates or non-experts. Prominent
examples include the lower dimensional mechanical 
analogs of massive Klein--Gordon theory
(harmonic oscillator), the Higgs model (anharmonic oscillator/double well),
sine Gordon theory (particle on a circle/periodic potential) and many
others. With this is mind, we are tempted to ask the question --what
is the zero dimensional/mechanical 
analog of the field theory of tachyon matter?

To understand this let us rewrite the action in zero dimensions obtained 
via the correspondence :-  $x^{i} \rightarrow t$, $T\rightarrow x$ and 
$V(T)\rightarrow V(x)$ :

\begin{equation}
S_0 = -\int dt V(x) \sqrt{1-{\dot x}^2}
\end{equation}

The equation of motion for this action (with the assumption that $1-{\dot x}^2
\neq 0$--the equality corresponds, as evident from the discussion below, of
the particle reaching a terminal velocity) gives :

\begin{equation}
\ddot x + f(x){\dot x}^2 = f(x)
\end{equation}

where $f(x) = -\frac{1}{V}\frac{\partial V}{\partial x}$.
Using the zero dimensional version of the tachyon potential, i.e. 
$V(x) = e^{-\alpha x}$ (we do away with the $\frac{1}{2}$ factor in
the potential for the field theory version)
we obtain the equation of motion :

\begin{equation}
\ddot x +\alpha {\dot x}^2 = \alpha
\end{equation}

Redefining $\dot x = \gamma \dot y$ we obtain :

\begin{equation} 
\ddot y +\alpha \gamma {\dot y}^2 = \frac{\alpha}{\gamma}
\end{equation}

The above equation is familiar to all of us.
Replacing $\alpha \gamma =\frac{\beta}{m}$ and $\frac{\alpha}{\gamma} = g$
we get back :

\begin{equation}
m\ddot y + \beta {\dot y}^2 = mg
\end{equation}

which is the equation of motion of a particle of mass $m$ moving in a
Newtonian gravitational field in the presence of quadratic damping. One can
therefore get back this equation from the action :

\begin{equation}
S_0 = -\int dt e^{-\frac{\beta y}{m}} \sqrt{1-\frac{\beta}{mg}{\dot y}^2}
\end{equation}

This Lagrangian is ofcourse in a dimensionless form. One may scale the
coordinate and multiply by appropriate factors to have an action with
the required dimensions. This is a straightforward exercise.
Thus the zero dimensional version of the field theory of tachyon matter
provides an action integral formulation of the motion under Newtonian
gravity  in the presence of quadratic damping.

One might ask -- is this the unique action which gives the above equation of
motion? To answer it, we look for a counterexample. Raising the Lagrangian
to its $n$th power (recall the fact that the square root and the squared
action for the motion of relativistic, massive test particles in a 
background gravitational
field give the same equations of motion)  we write an action of the form :

\begin{equation}
S_n = -\int dt e^{-\frac{n\beta y}{m}} \left (1-\frac{\beta}{mg}{\dot y}^2
\right )^{\frac{n}{2}}
\end{equation}

From the equations of motion for this action we find that it is only 
for $n=1$ (and no other value of $n$)we get the
equation for a particle moving in a Newtonian gravitational field in 
the presence of quadratic damping. Another possibility is replacing 
$\sqrt{1-{\dot x}^2}$ by $\sqrt{1+{\dot x}^2}$. This yields a wrong sign 
in the right hand side $m\ddot y +\beta {\dot x}^2 = -mg$.
It is true that there may be other actions which yield the
same equation of motion, but, for the moment, we prefer to remain
satisfied with the one above.

Though the equations of motion and solutions for the above problem are well
known we are not entirely sure if the 
action integral reformulation exists. Moreover, the connection
with the field theory of tachyon matter is new and perhaps worth mention.
Additionally, this action integral formulation could be useful if one
wishes to quantise the theory in the language of path integrals. The
task is quite formidable in view of the exponential and the square root
in the action. However, it is possible to expand the square root and
exponential for small $\beta$ and arrive at an action of the form (for
small $\beta$) :

\begin{equation}
S_0 \sim -\int dt \left (1-\frac{\beta y}{m} -\frac{\beta}{mg}{\dot y}^2\right )
\end{equation}

For the above action it is easy to write out the path integral following
the standard procedure for quadratic actions.  
Quadratic  damping can be treated as a perturbation over classical solutions
of $m\ddot y = mg$ and its effect on the
kernel can be ascertained through its $\beta$ dependence.
 
We also note that the Hamiltonian arising from the  
Lagrangian in (Eqn (7)) is given as: 

\begin{equation}
H = \sqrt{\frac{mg}{\beta}} \left (p^2 + \frac{\beta}{mg}e^{-\frac{2\beta y}{m}}\right )^{\frac{1}{2}} 
\end{equation}

The overall negative sign in the action given in 
(1) or (7) is crucial in order to
have a positive definite Hamiltonian.
Using the above one might attempt at deriving a quantum mechanical
or microscopic reason behind  quadratic damping. 
 
In addition, the generality of the approach manifest through the
freedom in choosing $V(x)$ may be useful in situations with more complicated
, spatially dependent, but quadratic in velocity, damping forces. Extensions to
higher (two or three) dimensions in space is possible though
the equations of motion there are no longer uncoupled -- the presence
of the nonlinear damping term being the cause behind this
feature.

It is tempting, as a follow-up exercise, to extend the above ideas to $1+1$
dimensional field theories, in particular, nonlinear field theories.
For static solutions of a nonlinear scalar field theory, the 
equations of motion can be written down by a straightforward mapping
of the variables of the mechanical system in Euclidean time to the 
field theoretic variables.
It is obvious that the 
${\phi'}^2$ term, which will be present in the equations of motion,
will damp the solutions in a way similar to the mechanical system
discussed above. However, even though static solutions for the field system
can indeed be written down, it is not clear to us what they actually mean
in the context of a field theory. 
These and similar issues provide avenues of further study.

Finally, it is a rather pleasant surprise that a field theory arising out
of such an involved and complicated enterprise as string field theory
does have a connection (albeit through an analogy) with a system which
we all know about and also experience in our everyday life. This, like
other theories with similar analogs, makes the field theory of tachyon matter
perhaps somewhat closer to reality. As mentioned in the beginning of this
article, what we have discussed here is just a zero dimensional analog--
its input in the actual field theoretic context is, to us, at this moment,
largely vacuous.   

The author thanks S. P. Khastgir and P. Majumdar for comments and discussion.

\end{document}